\documentclass[showpacs,floatfix,twocolumn,amssymb,amsfonts,amsmath,aps,pre]{revtex4}
\usepackage{graphicx,epsfig}

\usepackage{latexsym}
\usepackage{graphicx}
\usepackage{gensymb}

\usepackage{hyperref}

\newcommand{\dC}{$^{\circ}$C}
\newcommand{\etal}{\textit{et al.}}

\begin{document}
\title{Capillary Rise of Water in Hydrophilic Nanopores}
\author{Simon Gruener$^{1}$}
\email[]{s.gruener@mx.uni-saarland.de}
\author{Tommy Hofmann$^1$}
\author{Dirk Wallacher$^2$}
\author{Andriy V. Kityk$^3$}
\author{Patrick Huber$^{1}$}
\email[]{p.huber@physik.uni-saarland.de}
\affiliation{$^1$Faculty of Physics and Mechatronics Engineering, Saarland University,
D-66041 Saarbr\"ucken, Germany\\
$^2$Helmholtz Center for Materials and Energy, Glienicker Str. 100, D-14109 Berlin, Germany\\
$^3$Institute for Computer Science, Czestochowa University of Technology, Aleja Armii Krajowej 17, PL-42220 Czestochowa, Poland}


\begin{abstract}
We report on the capillary rise of water in three-dimensional networks of hydrophilic silica pores with $3.5$\,nm and $5$\,nm mean radii respectively (porous Vycor monoliths). We find classical square root of time Lucas-Washburn laws for the imbibition dynamics over the entire capillary rise times of up to 16\,h investigated. Provided we assume two preadsorbed, strongly bound layers of water molecules resting at the silica walls, which corresponds to a negative velocity slip length of -0.5\,nm for water flow in silica nanopores, we can describe the filling process by a retained fluidity and capillarity of water in the pore center. This anticipated partitioning in two dynamic components reflects the structural-thermodynamic partitioning in strongly silica bound water layers and capillary condensed water in the pore center which is documented by sorption isotherm measurements.
\end{abstract}



\maketitle

Water transport in nanoscale environments plays a crucial role for phenomena ranging from clay swelling, frost heave, and oil recovery to colloidal stability, protein folding, and transport in cells and tissues \cite{Relevance1}. The mobility of water has, therefore, been extensively studied by experiment and theory in such restricted geometries over the last three decades.\\
These studies have revealed a remarkable fluidity of water down to nanometer and even subnanometer spatial confinement \cite{WaterNanoFluidity} and have also demonstrated the validity of macroscopic capillarity concepts at the nanoscale and mesoscale \cite{WaterCapillarity}. However, how capillary forces along with the retained fluidity of water and other liquids contribute to the huge variety of phenomena, where self-propelled fluid transport is encountered in nanoscale geometries has been almost solely investigated so far by theory or for simple geometries as thin films \cite{Gelb2002}.\\
Here, we report an experimental study on the spontaneous imbibition, that is, the capillary rise (CR) of water in networks of hydrophilic silica pores with characteristic radii of $\mathcal{O}$(10\,nm). Our study, which extends a former more qualitative investigation of this phenomenology \cite{Huber2007}, is aimed at a quantitative description of this process in terms of the driving capillary forces, the fluidity of water, and the complex flow topology encountered. We shall also demonstrate that an understanding of the CR dynamics requires a detailed knowledge of the thermodynamic state of water in this nanoscale environment, which we achieved by sorption isotherm measurements.\\
\begin{figure} \center
\includegraphics[width=8.3cm]{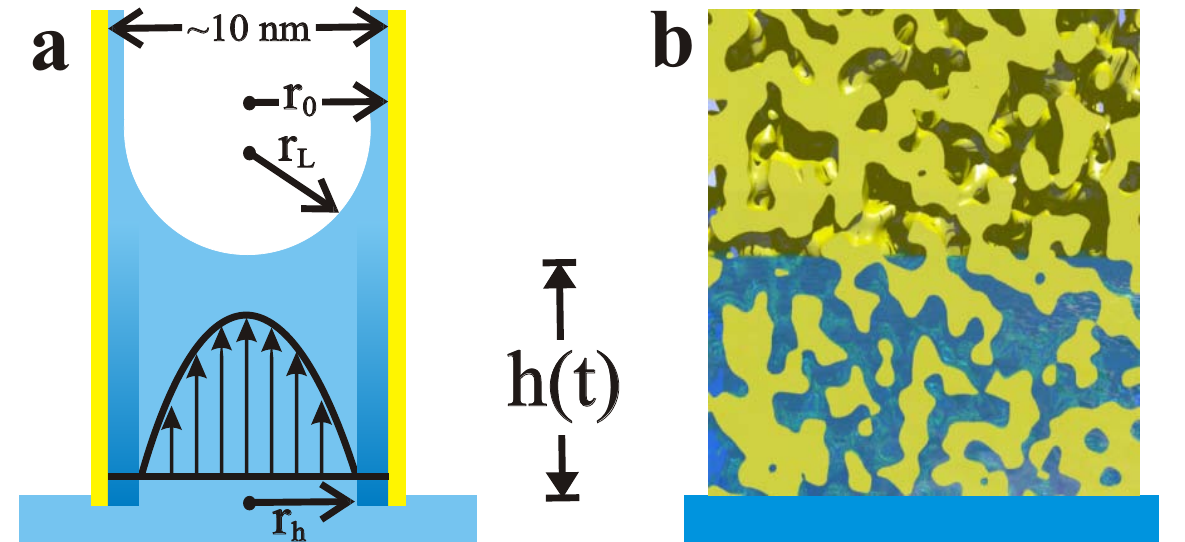}
\caption{\label{fig1}(Color online). (a) Schematic side view on the capillary rise of water in comparison with a raytracing illustration of spontaneous water imbibition in Vycor (b). A liquid column has advanced up to the height $h(t)$. A parabolic fluid velocity profile along with preadsorbed water layers beyond $h(t)$ and a shaded resting boundary layer are sketched for the nanocapillary in panel (a).}
\end{figure}
As porous host we have chosen "thirsty" Vycor glass (Corning glass, code 7930) \cite{Levitz1991}, a virtually pure fused silica glass permeated by a three-dimensional network of interconnected tortuous pores (see Fig. \ref{fig1}). The experiments were performed with two types of Vycor significantly differing only in the mean pore radius (3.4\,nm and 4.9\,nm respectively). We cut blocks of $\sim 30$\,mm height from the rods. Prior to using, we subjected them to a cleaning procedure with hydrogen peroxide and nitric acid followed by rinsing in deionized Millipore water and drying at 200\,\dC\, in vacuum for 2 days. Infrared spectroscopy then reveals that the silica surfaces are covered by silanol groups with $\sim 4$ {Si-OH}/nm$^{\rm 2}$ surface density  \cite{Wallacher1998}, rendering them hydrophilic with a contact angle for water droplets close to 0 \cite{Schulten2006}. At last, all facets but the bottom one were sealed. Water and nitrogen sorption experiments were performed with crunched Vycor samples at 4\,\dC\, and -196\,\dC, respectively, using an all-metal gas handling setup.\\
The mass uptake of the porous host as a function of imbibition time $m(t)$ was recorded gravimetrically with an electronic balance after immersing the monolith into a water bath thermostated to $T=24$\,\dC. Right upon contact of the bottom facet of the sample with the surface of the bulk liquid, we observed a jump in $m(t)$ of about $0.3$~g. It can be attributed to the formation of a macroscopic liquid meniscus at the perimeter of the matrix. Since we are interested in the water immersion into the inner pore space, we will present data sets after subtracting this mass jump. In principle, one has also to worry about buoyancy forces acting on the porous host and inertia effects in the initial states of the imbibition process \cite{Quere1997}. However, both contributions are negligible in our experiment; a statement supported by the data presented below.\\
As can be seen in Fig. \ref{fig2}a, we find a monotonically increasing $m(t)$ up to $t\sim 7$\,h and $16$\,h, respectively. At that time, the advancing water front reaches the top of the monolith, as can be observed by the eyes due to a small contrast in transparency between the empty and filled parts of the sample. The time behavior of the imbibition of water into the porous monolith is well known and follows a $\sqrt{t}$ Lucas-Washburn law \cite{Lucas1918}, which results from a competition of time invariant capillary forces driving the flow and an increasing viscous drag due to the increasing length of the paths connecting the advancing water front with the bulk reservoir. For macroscopic capillaries, such a simple scaling is only observable in the initial states of an imbibition process, eventually gravity comes into play leading to an exponential relaxation toward a final CR height characterized then by a balance of capillarity and gravity. By contrast, for liquids, wetting the inner surfaces of capillaries with radii $r_{\rm L}$ of a couple of nanometers, the driving capillary pressure, given by Laplace's law,
\begin{equation}
\label{EqLaplace}
p_{\rm L} = \frac{2 \, \sigma }{r_{\rm L}},
\end{equation}
can be on the order of several hundred bars. In particular, for water with its surface tension $\sigma$ of 72\,mN/m, Eq. \ref{EqLaplace} yields a $p_{\rm L}\sim 300$\,bar for a pore with radius $r_{\rm L}=5$\,nm. This corresponds to a hypothetical CR height of 3\,km \cite{Caupin2008}. Thus, gravity forces should be negligible in our experiment and, accordingly, we find in accord with most recent microscopic simulations for liquid imbibition into nanopores \cite{Gelb2002} an excellent agreement with the Lucas-Washburn prediction over the entire $t$-range investigated (see the minute residuals shown in Fig. \ref{fig2}b).\\
\begin{figure} \center
\includegraphics[width=7.5cm]{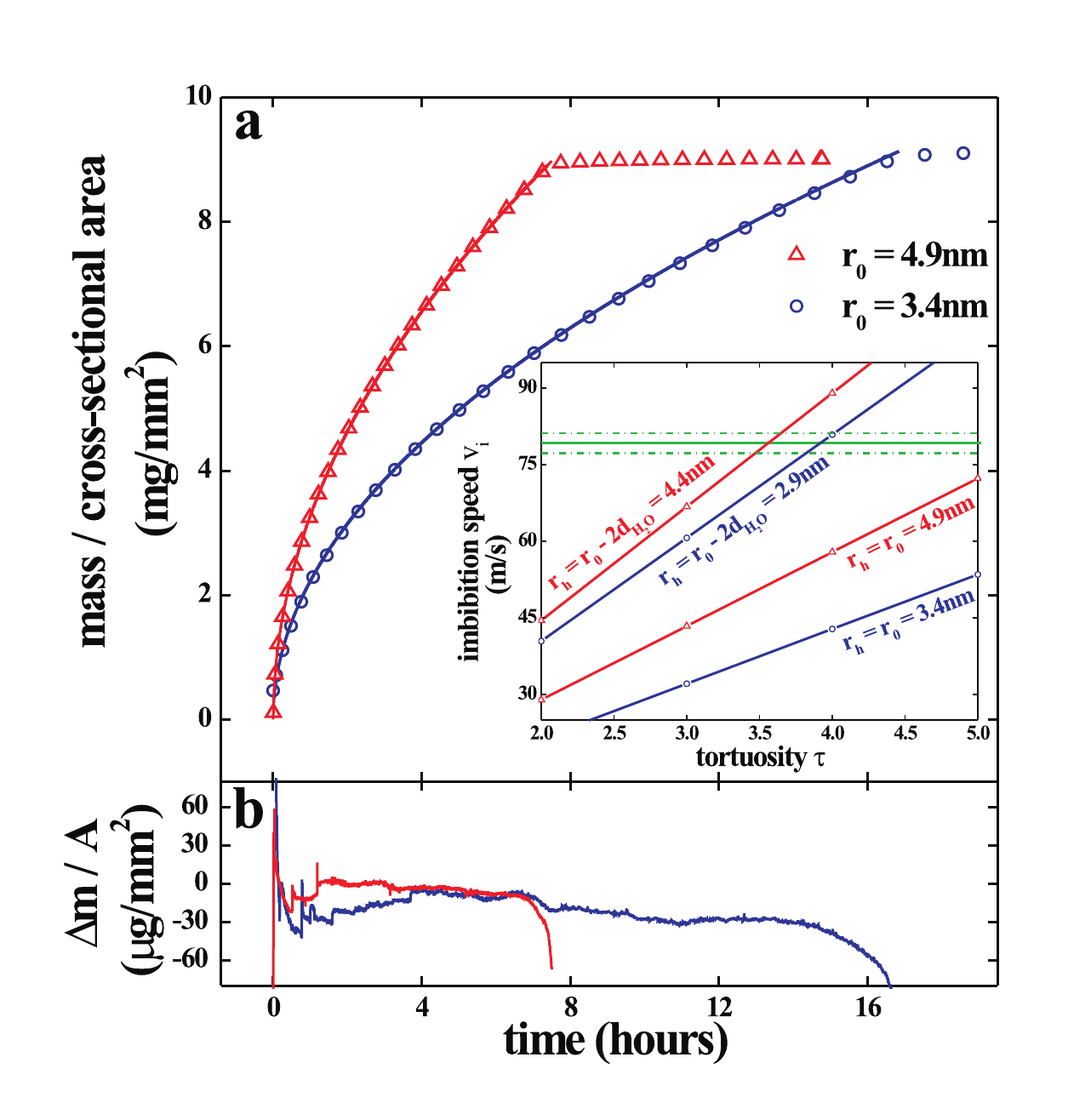}
\caption{\label{fig2}(Color online). (a) Normalized mass uptake $m(t)$ of Vycor with two different pore radii ($r_0=4.9$\,nm: triangles, $r_0=3.4$\,nm: circles) due to water imbibition as a function of time $t$ in comparison with Lucas-Washburn $\sqrt{t}$-fits (solid lines). Inset: imbibition speed $v_{\rm i}$ - tortuosity $\tau$ map. The lines represent CR speeds determined with Eq. \ref{Eqmt} from the measured $m(t)$ rate as a function of the tortuosity $\tau$ and for two different hydraulic pore radii ($r_{\rm h}=r_0$ and $r_{\rm h}=r_0-2\, d$, respectively, where $d=0.25$~nm denotes the diameter of a water molecule) for both types of Vycor as indicated. The predicted $v_{\rm i}$ value for bulk water and its error margins are represented by horizontal lines. (b) Residuals $\Delta m$ between the measured $m(t)$-curve and the $\sqrt{t}$-fits.}
\end{figure}
More detailed quantitative insights regarding our CR experiment can be achieved by referring to the physics of liquid flow in porous media. According to Darcy's law, the volume flow rate $\dot V$ through a surface area $A$ of a porous host which is subjected to a hydrostatic pressure $\Delta p$ applied along the height $h$ is given by
\begin{equation}
\label{EqDarcy}
\frac{\dot V}{A} = \frac{K}{\eta}\, \frac{\Delta p}{h},
\end{equation}
where $\eta$ is the shear viscosity of the liquid, and $K$ refers to the hydraulic permeability of the porous matrix.\\
For many porous systems with reasonable homogeneous cylindrical pore structures, among them also for Vycor, it has been demonstrated that $K$ can be well related to phenomenological quantities typical of the matrix's morphology \cite{Lin1992,Debye1959},
\begin{equation}
\label{EqPermeab}
K=\frac{1}{8}\, \frac{r_{\rm h}^{\rm 4}\, \phi_0}{r_0^2\,\tau},
\end{equation}
where $\phi_0$ is the volume porosity of the host, and $\tau$ is its tortuosity characterizing the connectivity and meandering of the pores. Note that the hydrodynamic radius $r_{\rm h}$, which is the radius over which a parabolic flow profile is established (see Fig. \ref{fig1}a), does not necessarily have to agree with the mean pore radius $r_0$ because of either strongly adsorbed, immobile boundary layers, or due to velocity slippage at the pore walls \cite{slip, StoneBrenner}. In a CR geometry, the liquid encounters the pressure drop along a time-dependent length $h(t)$ (see Fig. \ref{fig1}b), and $\Delta p$ is solely determined by the Laplace pressure $\Delta p=p_{\rm L}$ that acts at the advancing water menisci. Moreover, with the initial porosity $\phi_{\rm i}$ of the sample upon CR start the permeated volume can be determined to be $V(t)=\phi_{\rm i} \,A\, h(t)$. Accordingly, Darcy's law transforms to a simple differential equation,
\begin{equation}
\label{EqDiff}
\dot V \,V= A^2\,\frac{K}{\eta}\, p_{\rm L}\, \phi_{\rm i}.
\end{equation}
Equation \ref{EqDiff}~is solved by a $\sqrt{t}$ law for $V$, which by multiplication with the mass density $\rho$ of the liquid yields the aforementioned Lucas-Washburn law for the mass uptake,
\begin{equation}
\label{Eqmt}
m^2(t)= \rho^2\,A^2\,\underbrace{\frac{r_{\rm h}^4\,\phi_0\,\phi_{\rm i}}{2\, r_0^2 \, r_{\rm L}\,\tau}}_{G} \underbrace{ \frac{\sigma}{\eta} }_{v_{\rm i}} t.
\end{equation}
As can be seen in Eq. \ref{Eqmt}, the ratio of surface tension to viscosity of pore-confined water, to which we refer in the following as imbibition speed $v_{\rm i}$, is the property of the liquid which determines the dynamics of the CR. Thus, we can determine this ratio for confined water from our CR experiment provided we can assess the geometry factor $G$ of our porous host.\\
The parameters $\phi_0$ and $r_0$ are accessible by means of a gas sorption isotherms, which motivated us to record nitrogen and water vapor isotherms at $T=-196$\,\dC\, and $T=4$\,\dC, respectively. The latter is presented in Fig. \ref{fig3}. We plot the filling fraction $f$, that is, the number of water molecules adsorbed by the matrix normalized to the water amount necessary for its complete filling, versus the reduced vapor pressure $p=p_{\rm e}/p_{\rm 0}$. The pressure $p_{\rm 0}$ refers to the bulk vapor pressure of water at $T=4$\,\dC\, and $p_{\rm e}$ to the equilibration pressure after each adsorption or desorption step, respectively. The initial steep increase of $f$ with $p$ up to point A indicated in Fig. \ref{fig3} is due to the formation of a first layer of water molecules which is strongly bound to the silica surface and hence exhibits a small $p$. Along the path A$\rightarrow$B, a second layer grows. After the formation of the third layer in point C ($f\approx 0.3$), we enter the hysteretic capillary condensation/evaporation regime \cite{Huber1999}, [(C)-(E)]. The entire pore space fills or empties via the formation or retreat of capillary bridges (see the inset in Fig. \ref{fig3}). At point F, the matrix is completely filled and we measure $p_{\rm 0}$ typical of bulk droplets sitting outside the pores. The overall number of adsorbed molecules yields a porosity of $ (31\pm 2)$\%. The mean pore diameters were determined from an analysis of the nitrogen sorption isotherms to ($6.8\pm 0.2$)\,nm and ($9.8\pm 0.2$)\,nm based on a mean-field model for capillary condensation proposed by Saam and Cole \cite{Huber1999}. It is important to note that, according to the isotherms, the final relative humidity in our laboratory of $\sim$ 30\% ($p=0.3$) will lead to at least one layer of strongly adsorbed water. This also means that the initial porosity $\phi_{\rm i}$ in our CR experiments is reduced by $\sim 10$\%. Additionally, we have to assume $r_{\rm L}$ to be reduced in comparison to $r_0$ by the thickness of the preadsorbed water film.\\
\begin{figure} \center
\includegraphics[width=7.5cm]{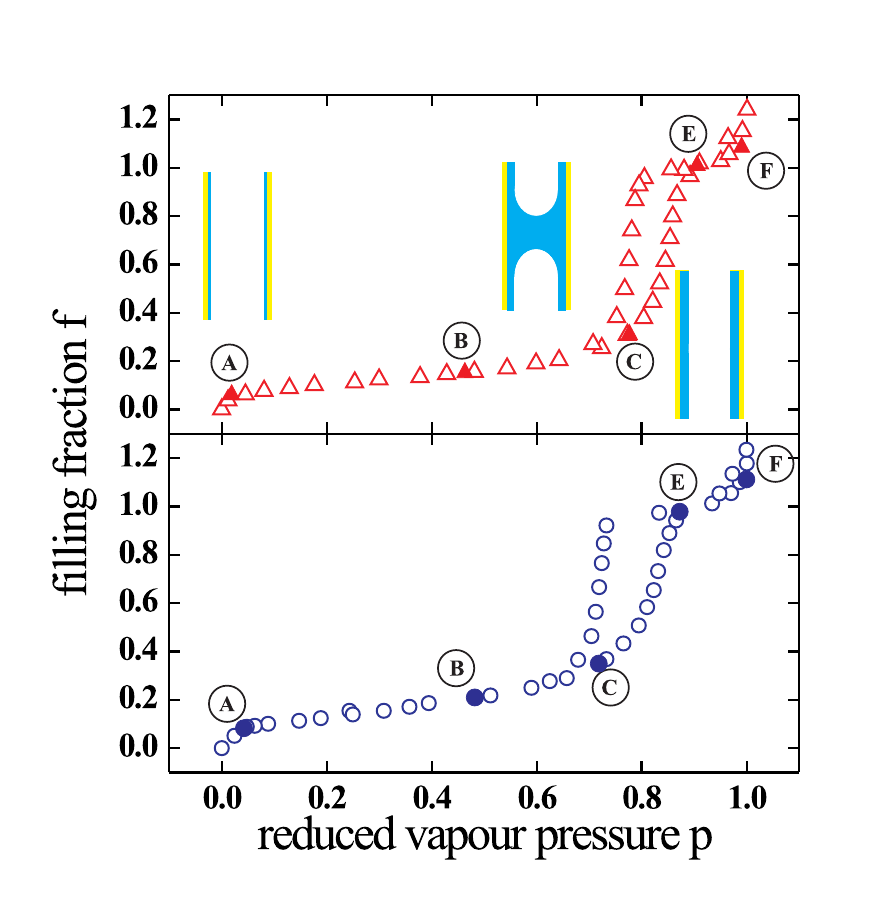}
\caption{\label{fig3} (Color online). Water isotherms recorded at $T=4$\,\dC\, for Vycor with 9.8\,nm (triangles) and 6.8\,nm (circles) mean pore diameter. Plotted is the filling fraction $f$ versus the reduced vapor pressure $p$. Five characteristic points discussed in the text are indicated by solid symbols.}
\end{figure}
The tortuosity could be inferred from experiments on the self-diffusion of liquids in porous Vycor \cite{Lin1992} to $\tau= 3.6 \pm 0.4$, in accordance with simulations of its pore morphology \cite{Crossley1991}. A value of $\tau$ in the proximity of three seems reasonable if one considers that in an isotropic spongelike medium such as Vycor, the porosity can, in first approximation, be accounted for by three sets of parallel capillaries in the three spatial directions; whereas only 1/3 of these capillaries sustains the flow along the pressure drop $\Delta p$. A value larger than three reflects then the extended length of a meandering capillary beyond that of a straight one.\\
We can now determine the CR speed $v_{\rm i}$ from our mass uptake measurements. In order to allow the reader a judgement on the influence of the uncertainties in $\tau$ and $r_{\rm h}$ on $v_{\rm i}$, we plot the derived imbibition speeds as a function of $\tau$ for selected assumed hydraulic radii $r_{\rm h}$ (see the inset in Fig. \ref{fig2}a). In principle, all of the resulting lines run in the vicinity of $v_{\rm i}=(79\pm 2)\,\frac{{\rm m}}{{\rm s}}$ calculated from bulk viscosity and surface tension of water at $T=24$\,\dC\ and indicated by the horizontal lines. More interestingly, for $\tau=3.6\pm0.4$, both experiments reproduce the predicted bulk value for $v_{\rm i}$ ($v_{\rm i}=(73\pm 40)\,\frac{{\rm m}}{{\rm s}}$ and $v_{\rm i}=(80\pm 33)\,\frac{{\rm m}}{{\rm s}}$ for Vycor with $r_0=3.4$\,nm and $r_0=4.9$\,nm, respectively) provided we assume two monolayers of water adjacent to the pore walls to be immobile ($r_{\rm h}=r_0-2\,d$, where $d=0.25$~nm refers to the diameter of a water molecule). Note, for $r_{\rm h}=r_0$, we systematically underestimate $v_{\rm i}$ in both cases. Thus, our experiments confirm former findings on the conserved fluidity \cite{WaterNanoFluidity} and capillarity \cite{WaterCapillarity} of nanoconfined water, if we assume a velocity slip length; that is, the distance with respect to the fluid/solid interface, where the flow velocity vanishes, of $-0.5$\,nm. This assumption is corroborated by recent molecular-dynamics studies on the glassy structure of water boundary layers in Vycor and the expected existence of sticky boundary layers in Hagen-Poiseuille nanochannel flows for strong fluid-wall interactions \cite{Gallo2001}. Also tip-surface measurements document a sudden increase of the viscosity by orders of magnitude in 0.5\,nm proximity to hydrophilic glass surfaces \cite{Li2007}. It also extends former experimental findings with respect to the validity of the no-slip boundary condition for water/silica interfaces, where this condition has been proven down to at least 10~nm from the surface \cite{Lasne2008}. Moreover, the retained bulk behavior agrees with our observation of a correct relative scaling of measured and expected $v_{\rm i}$ values for water and a set of n-alkanes upon invasion in nanopores\cite{Huber2007}.\\
Our experiments yield additional rheological details. As mentioned above, Eq.~\ref{EqPermeab} intrinsically assumes a parabolic flow velocity profile across the pore cross-section, which implies a linear variation in the viscous shear rate, starting with 0 in the pore center to a $t$-dependent maximum at $r_{\rm h}$, $\dot{\gamma}_{\rm m} \propto \frac{1}{\sqrt{t}}$. For example, for the $r_{\rm h}=2.9$\,nm experiment, we estimate $\dot{\gamma}_{\rm m}$ to decrease from $7\times 10^4\,\frac{1}{\rm s}$ after 1\,s to $3\times 10^2\,\frac{1}{\rm s}$ at the end of the CR. As we found no $t$-dependent and, therefore, no $\dot{\gamma}$-dependent deviations of $m(t)$ from a single $\sqrt{t}$-fit, our measurements also testify the absence of any non-Newtonian behavior of water as well as an unchanged no-slip boundary condition imposed at $r_{\rm h}$. Despite the relatively large $\dot{\gamma}$'s probed here, the latter is not too surprising. The viscous forces of $\mathcal{O}(\eta\,d^2\, \dot{\gamma})$ can only overcome the strong water/silica interactions of $\mathcal{O}(A/d)$ (Hamaker constant $A\sim 10^{-19}$\,J) for $\dot{\gamma} > 10^{12}\,\frac{1}{\rm s}$ \cite{StoneBrenner, Tabeling2004} -- significantly beyond the $\dot{\gamma}$'s probed here.\\
It is worthwhile to note that water encounters a negative pressure upon CR which linearly decreases from $-p_{\rm L}$ at the advancing menisci to atmospheric pressure at the bulk reservoir. Water's hydrogen bridge bond network is expected to be responsible for an increase in $\eta$ and decrease in $\rho$ under such large tensile pressures. Based on thermodynamic models for stretched water \cite{Stanley2002}, we estimated an $\sim 3 \%$ $v_{\rm i}$ decrease due to this effect, which is, unfortunately, well below our error margins in $v_{\rm i}$.\\
In summary, we have demonstrated classical Lucas-Washburn CR dynamics for water in networks of hydrophilic silica nanopores by rather simple gravimetric imbibition experiments. They can be described in reasonable way by macroscopic hydrodynamics provided a sticky preadsorbed boundary layer of about two monolayers of water molecules is considered. We hope that the CR study presented here will stimulate further experiments on the interplay of fluidity and capillarity on the nanoscale not only for water but also for other complex liquids \cite{Eijkel2005}. In particular, we envision experiments on carbon nanotubes with graphitic less water attractive walls, where single file flow and velocity slippage may affect the CR dynamics markedly \cite{Majumder2005}.\\

\begin{acknowledgments}
We thank K. Knorr, H. Stone, and H. Tanaka for helpful comments and acknowledge support within the DFG priority program 1164, \textit{Nano- and Microfluidics}, Grant No. Hu 850/2.
\end{acknowledgments}


\end{document}